\documentstyle[epsfig,emulateapj]{article}

\lefthead{Meier}
\righthead{MAGNETICALLY-SWITCHED, ROTATING BLACK HOLE JET MODEL}

\def\bm#1{\mbox{\boldmath{$#1$}}}

\def\sm{{\scriptscriptstyle{-}}}

\received{ }

\slugcomment{Submitted to The Astrophysical Journal.}

\begin{document}

\hyphenation{ }

\title{A Magnetically-Switched, Rotating Black Hole Model \\
    For the Production of Extragalactic Radio Jets and the \\
    Fanaroff and Riley Class Division}

\author{D. L. Meier}
\affil{Jet Propulsion Laboratory, California Institute of Technology,
    Pasadena, CA 91109}




\begin{abstract}
A model is presented in which both Fanaroff and Riley class I and II 
extragalactic jets are produced by magnetized accretion disk coronae 
in the ergospheres of rotating black holes.
It employs a hybrid version of the Blandford-Payne and Blandford-Znajek 
magnetohydrodynamic mechanisms (similar to the Punsly-Coroniti model, with 
the addition of a metric-shear-driven dynamo) and a generalized form of the magnetic 
switch, which is shown to be the MHD analog of the Eddington luminosity.  
While the jets are produced in the ergospheric accretion disk 
itself, the output power still is an increasing function of the black hole 
angular momentum.  For high enough spin, the black hole triggers the 
magnetic switch, producing highly-relativistic, kinetic-energy-dominated 
jets instead of magnetic-energy-dominated ones for lower spin.  The 
coronal mass densities needed to trigger the switch at the observed 
FR break power are quite small ($\sim 10^{-15} \, {\rm g \, cm^{-3}}$), 
implying that the source of the jet material may be either a pair plasma or 
very tenuous electron-proton corona, not the main accretion disk itself.

The model explains the differences in morphology and Mach number 
between FR I and II sources and the observed trend for massive galaxies 
(which contain more massive black holes) to undergo the FR I/II 
transition at higher radio power.  It also is 
consistent with the energy content of extended radio lobes and explains 
why, because of black hole spindown, the space density of FR II sources 
should evolve more rapidly than that of FR I sources. 

A specific observational test is proposed to distinguish between models 
like this one, in which the FR I/II division arises from processes near 
the black hole, and models like Bicknell's, in which the difference 
is produced by processes in the host galaxy's interstellar medium. If the 
present model is correct, then the ensemble average speed of parsec-scale jets 
in sources distinguished by their FR I {\em morphology} (not luminosity) 
should be distinctly slower than that for sources with FR II morphology.
The model also suggests the existence of a population of high-redshift, 
sub-mJy FR I and II radio sources associated with spiral 
or pre-spiral galaxies that flared once when their black holes were 
formed but were never again re-kindled by mergers. 
\end{abstract}


%
%


\section{Introduction}

Recently many authors have pointed out that the speeds of jets associated with 
different types of compact object (pre-main-sequence star, white dwarf, 
neutron star, black hole) are approximately equal to the Keplerian or 
escape speed from the surfaces of those compact objects.  
However, some caution must be exercised in taking this equivalence 
too literally.  A good example of this is the extragalactic radio source 
class.  All are believed to be powered by accretion onto black holes, from 
whose surface (the horizon) the escape velocity is exactly $c$, the speed of 
light.  However, while all extragalactic jets appear to be at least mildly relativistic, 
the speeds obtained from Very Long Baseline Interferometry (VLBI) observations 
range from $\beta_{j} \equiv v_{j}/c \sim 0.1$ in Centaurus A (\cite{ting98}) to 
$\gamma_{j} \equiv (1 - \beta_{j}^2)^{-1/2} \sim 10$ in 3C 345 (\cite{verm96}).  
In terms of the proper velocity 
$u_{j} = \gamma_{j} \beta_{j}$, this represents a range of two orders 
of magnitude in jet speed.  Clearly, even before traversing the main part of 
the interstellar medium, jet speeds are strongly affected by the details 
of the acceleration and collimation process near or in the accreting black 
hole system.

Radio jets on the kiloparsec scale also differ dramatically in speed, and 
these differences apparently translate into a morphological difference.  
Subsonic or transonic jets with velocities less than $\sim 0.6 c$ are 
believed to be responsible for producing the Fanaroff and Riley Class I sources 
(\cite{fr74};  \cite{bick85}; \cite{bick95}).  The FR I sources are distinguished by 
generally low radio power, often 
complex morphology (S-shaped, C-shaped, {\it etc.}), and diffuse lobes that 
have their bright region close to the associated galaxy --- all consistent 
with transonic flow that decelerates into subsonic flow.  Class II sources 
generally are much more powerful, have only a straight morphology, and have their 
brightest point at the far ends of the dumbbell-shaped source.  They are 
thought to be produced by highly supersonic jets that terminate in a 
strong shock where the high-speed flow strikes the interstellar medium.  

A mechanism for determining the speed of jets at their site of acceleration 
was suggested by \cite{m97a}, hereinafter MEGPL.
Termed the ``magnetic switch'' because of the two distinct slow and fast 
jet states, this mechanism operates in the magnetized corona of the black hole 
accretion disk itself.  Fast (highly relativistic) jets are produced when the 
coronal plasma is unbound by the magnetic field, and slow (trans-relativistic) 
jets at roughly the disk escape velocity are produced when the corona is 
initially bound.  This simple physical process is a good candidate for 
explaining the difference between the two Fanaroff \& Riley classes, because 
it explains how jets with radically different speeds could occur in otherwise 
similar sources.  However, as discussed in section \ref{old_magsw_section} 
below, the magnetic switch, triggered solely by the magnetic field strength 
(or even by coronal density), fails to explain why these morphological differences 
have rather long lifetimes ($10^{6-7}$ yr).    

One way out of this dilemma is to assume that all radio jets for a given 
galaxy mass indeed do start out in the nucleus with about the same average 
speed, but, depending on local galaxy properties (which do have long lifetimes), are affected 
differently by deceleration processes in the interstellar medium (\cite{bick95}).
Indeed, jet deceleration {\em has} been inferred from variations in the 
jet-to-counterjet ratio in observations of FR I sources (\cite{laing96}).  In this 
paper, however, we propose an alternate method for maintaining long-lived FR I 
or FR II morphology --- one that preserves the magnetic switch model, 
along with all of its attractive properties, and suggests a reason {\em why} 
some jets may be more prone to deceleration than others. Since the power in 
these MHD outflows also depends on the angular velocity of the magnetic field, 
{\em it is suggested that the difference in radio power among galaxies of 
the same mass is due to different magnetic field rotation rates in their 
central engine, caused by different spin rates of their central black holes}.  
This has been suggested before (\cite{baum95}) based on observations alone, but 
this paper derives such a behavior directly from a generalized theory of the 
magnetic switch.
As we show below, black holes of different spin should produce jets of 
different power and velocity, and a sharp FR I/II-like transition is 
expected to occur.  Unlike the ISM model, this model predicts a different 
mean jet velocity in the central parsec for the FR I and FR II classes, 
yielding a definite observational test for distinguishing between the two. 

Note that the magnetically-switched, rotating black hole model depends critically 
on the jet luminosity being a 
direct function of the black hole angular momentum --- in direct conflict 
with the conclusions of \cite{lop99}.  Therefore, section \ref{rot_bh_subsection}
spends some time on this issue and shows that, while the basic tenets of 
these authors are correct, nevertheless, if the accretion disk within the 
ergosphere is taken into account, a Blandford-Znajek-like (BZ) expression 
for the jet power is recovered --- even though the jet is ejected from the 
accretion disk itself by a Blandford-Payne-like (BP) mechanism.  The jet 
production model is similar to the Punsly-Coroniti (PC) ergospheric wind 
mechanism (\cite{pc90b}). Furthermore, due to an expected {\em metric-shear}-driven 
dynamo (see the Appendix), the poloidal magnetic field in the ergosphere is expected 
to have a larger value than that suggested by \cite{lop99}, leading to radio jet 
powers comparable to those observed.

The outline of the paper is as follows.  Section \ref{old_magsw_section} briefly 
reviews the classical magnetic switch model and the reasons for its failure to 
explain the FR I/II break.  Section \ref{gen_magsw_section} presents the generalized 
magnetic switch (including rotation as a parameter) and its role in the rotating black 
hole model for jet production and for the FR I/II transition.  Section \ref{appl_section} 
discusses applications of the model to recent observations of radio sources, with 
observational and theoretical tests of the model proposed in section \ref{tests_section}.

\section{The Magnetic Switch for Normal Compact Objects and Non-Rotating 
Black Holes}

\label{old_magsw_section}

\subsection{The Classical Magnetic Switch}

In the original magnetic switch model presented in MEGPL, the flow behavior 
depends on 
the ratio of the Alfv\'{e}n velocity in the inner disk corona $V_{A0} \equiv 
B_{p0}/(4 \pi \rho_{c0})^{1/2}$ to the escape velocity there $V_{esc0} = 
(2 G M/R_{0})^{1/2}$, where $B_{p0}$ is the strength of the poloidal 
magnetic field protruding from the disk into the corona at radius $R_{0}$, 
$\rho_{c0}$ is the density of the corona there, $M$ is the mass of the 
black hole, and $G$ is the gravitational constant.  When $\nu \equiv 
V_{A0}/V_{esc0} < 1$, gravity dominates magnetic forces (the plasma is bound 
initially) and the character 
of the flow is similar to a Parker wind (with collimation):  
azimuthal magnetic field coiled in the corona uncoils upward to slowly 
accelerate the flow through a series of critical points.  Depending on the 
parameters, the resulting jet speed is approximately equal to the escape 
speed at the base of the outflow (\cite{ks96}), which is somewhat interior 
to the last stable orbit in the relativistic Schwarzschild black hole case 
(\cite{koide98a}).  
When $\nu > 1$, magnetic forces dominate gravity and, for sufficiently high 
$\nu$, render the latter negligible.  (The coronal plasma is unbound.)  
As in pulsar wind theory, the 
outflow velocity then is determined not by gravitational effects, but by 
the magnetization parameter 
\begin{equation}
\label{sigma_eq}
\sigma_{0} \; \equiv \; V_{A0}^2 R_{0}^2 / (4 V_{i0} c R_{L}^2)
\end{equation}
which is a measure of the amount of work done on the injected plasma;  here, 
$V_{i0}$ is the velocity with which matter is injected, $R_{L} \equiv 
c/\Omega$ is the radius of the light cylinder, and $\Omega$ is the angular 
velocity of the poloidal magnetic field at $R_{0}$.  For relativistic flows, 
$\gamma_{j} \approx \sigma_{0}$ appears to be a good approximation for the 
final MHD wind/jet velocity (\cite{c89}; \cite{h89}).

A transition between 
the two types of outflow then is induced by increasing the magnetic field 
strength $B_{p0}$ until $\nu > 1$, whereupon significantly higher jet 
speeds occur.  This magnetic switch has several properties that 
make it an attractive model for the difference in morphology between 
class I and class II extragalactic radio sources (\cite{fr74}):  1) the 
transition is sharp, occurring with increasing jet power; 2) jets 
produced when $\nu < 1$ are transonic, while those for $\nu > 1$ are 
highly supersonic, 
and 3) even statistical behavior 
observed in large samples of radio galaxies of different power and optical 
magnitude is reproduced (see \cite{m97b} and Section 
\ref{olwl_subsection} below). 

\subsection{Failure of a Change in Magnetic Field Strength to Explain the 
FR I/II Break}

However, as presented in MEGPL, the original magnetic switch, triggered 
by either increasing the coronal magnetic field or decreasing the 
coronal density, fails to explain the FR I/II break, chiefly because it 
fails to explain why radio sources appear to maintain their FR morphology 
for long periods of time --- at least as long as the flow time from 
galactic center to lobe ($10^{6-7}$ yr).  There is no reason to expect 
the strength of the magnetic field threading disk and hole (or the coronal 
density) to remain fixed at high or at low values over such long time scales.
Indeed, one expects that these quantities will be distributed fairly 
equally about a mean, and on average will be the same for galaxies of the 
same mass.  The argument is as follows.  The structure of an accretion 
disk is a function of the black hole mass $m_{9} \equiv M \, / \, 10^9 
M_{\sun}$, the accretion rate $\dot{M}$, the viscosity parameter $\alpha$, 
and position in the disk scaled to the Schwarzschild radius (\cite{ss73}; 
\cite{nt73}). 
Now, there is growing evidence that the black hole mass, even in presently 
inactive objects, is approximately proportional to the mass of the central 
spheroidal bulge of stars (see, {\it e.g.}, \cite{kr95}) such that
\begin{equation}
\label{bh_mass_vs_gal_mag_eq}
\log m_{9} \; \approx \; -0.5 M_{B} \; - \; 10.5
\end{equation}
where $M_{B}$ is the absolute blue optical magnitude of the bulge.  
In addition, while the instantaneous accretion rate will vary dramatically 
on short time scales (years), because the hole is fed by galactic stars and 
gas, over the very long term the average accretion rate also should be 
roughly proportional to the bulge mass (a perhaps necessary condition for 
equation [\ref{bh_mass_vs_gal_mag_eq}] to be valid).  Therefore, to first 
order, we expect the average accretion rate 
\begin{equation}
\label{ave_mdot_eq}
<\dot{m}> \; \equiv \; <\dot{M}>/\dot{M}_{Edd} \; = <\dot{M}> / (1.4 \times 10^{26} \, 
{\rm g \, s^{-1}} \, m_{9})
\end{equation}
to be a constant, proportional to the average amount of matter shed per unit 
time into the nucleus and onto the hole by a galactic bulge of a given optical 
magnitude, divided by the mass of the black hole in that nucleus.  (Note that 
the definition of $\dot{m}$ here does not include an efficiency of conversion 
of accreting matter into radiation, allowing for maximum values somewhat greater 
than unity [$\dot{m}_{Edd} = 2.5 - 10$], depending on the spin of the black hole.  
It also assumes that the accretion is continuous; to take intermittent 
accretion into account, we will add a duty cycle parameter below.)  
Furthermore, detailed simulations of the magneto-rotational instability 
(\cite{stone96}; \cite{bnst96}) have shown that 
$\alpha \sim 10^{-2}$ is a fairly good approximation for turbulent accretion 
flows, presumably independent of black hole mass to first order.  And, finally, jets are 
believed to be ejected from near the black hole horizon ({\it i.e.}, from 
just inside the last stable orbit, \cite{koide98a}), rendering the position parameter 
$R_{0}/R_{Sch}$ a near-constant value for all objects.  Therefore, ignoring 
external influences, {\em over periods of time comparable to the flow time 
from nucleus to lobe, 
all parameters that affect the disk structure equations are expected to be, 
at most, functions of the black hole/galactic bulge mass only}.  On average, 
one expects bulges of the same mass to produce the same type of accretion 
disk in the nucleus, with similar magnetic field and coronal density, and, 
therefore, the same 
type of jet (either all slow or all fast). It is not expected that there will 
be a spectrum of jet structures that maintain themselves for millions of years.
Yet the observations show that just such a range of radio powers 
and morphologies exists, and that a radio source remains an FR I or II for 
longer than a flow time.

\section{The Magnetic Switch in the Rotating Black Hole Environment}

\label{gen_magsw_section}

In order to lift the degeneracy in expected jet speeds and morphologies, we 
shall take as an additional parameter the rotation rate of the magnetic field, 
as determined by the black hole rotation rate itself.  
This section develops the model of the accreting rotating black hole system, 
the production of radio jets, and the triggering of the magnetic switch.

In order to compute observable quantities, we must assume an accretion scenario 
in the black hole system.  For definiteness, we choose a rather benign (standard) accretion 
disk in which electron scattering dominates free-free absorption and gas 
pressure dominates radiation pressure, and we scale the results to a rather 
low accretion 
rate ($\dot{m} \sim 10^{-3}$).  While the details will change depending on the 
exact accretion model and rate, the basic conclusions herein will not, even in 
the extreme cases of near-Eddington accretion (radiation pressure dominated) or 
very low, advection-dominated accretion flow (ADAF; \cite{ny95}).

\subsection{The Generalized Magnetic Switch}

The first task is to generalize the magnetic switch theory to include rotation as 
a triggering parameter, in addition to magnetic field strength and plasma density.
Consider a general magnetic rotator of size $R_{0}$, 
poloidal magnetic field $B_{p0}$, and angular velocity $\Omega$.  Following 
\cite{bp82}, the work done on the plasma trapped in the field lines, resulting 
in an outflowing magnetohydrodynamic (MHD) wind or jet, is
\begin{equation}
\label{nonrel_mhd_power_eq}
L_{MHD} \; = \; B_{p0}^2 \, R_{0}^3 \, \Omega / 2
\end{equation}
for non-relativistic flow.  
Now, for a normal Parker-type wind, the plasma in the injection region is 
in hydrostatic equilibrium, with only a slight force imbalance outward that 
slowly accelerates the outflow through a critical point until it reaches 
escape velocity and leaves the system (see, {\it e.g.}, \cite{m82}).  This 
is also true of MHD winds, although the critical point structure and 
geometry are more complex.  However, based on the magnetic switch mechanism 
described by MEGPL, there should exist a critical power, given by the 
liberation of an escape energy in a free-fall time
\begin{equation}
\label{lcrit_eq}
L_{crit} \; = \; E_{esc0} / \tau_{ff0} \; = \; 4 \pi \rho_{c0} \, R_{0}^2 \, 
\left( \frac{GM}{R_{0}} \right)^{3/2}
\end{equation}
such that if $L_{MHD}$ exceeds $L_{crit}$, then the plasma becomes unbound 
--- even in the injection region itself.  That is, the outward magnetic and 
centrifugal forces there strongly exceed gravity, accelerating the plasma 
to greater than the escape velocity within $R_{0}$ and eventually to a much 
greater final velocity.  {\em The magnetic switch luminosity, then, plays the 
same role in MHD acceleration as the Eddington luminosity plays in radiative 
acceleration.} For Keplerian disks (with $\Omega \, = \, \Omega_{K} 
\, \equiv \, (GM/R_{0})^{3/2}$), the condition $L_{MHD} > L_{crit}$ reduces 
to the original magnetic switch condition $V_{A0} > V_{esc0}$. 
However, for a non-Keplerian rotator of {\em fixed} magnetic field 
strength and plasma density, there should exist a critical angular velocity 
\begin{equation}
\label{omega_crit_eq}
\Omega_{crit} \; = \; \Omega_{K} \, \frac{V_{esc0}^2}{V_{A0}^2}
\end{equation}
such that if $\Omega$ exceeds $\Omega_{crit}$, then the magnetic switch also 
should be triggered and a fast jet of the type found by MEGPL should ensue.

The author tested this hypothesis by performing numerical MHD simulations similar 
to MEGPL, but holding $B_{p0}$ fixed and varying the disk rotation speed 
instead.  The left panel of Figure \ref{sim_fig} shows the resulting jet speed 
as a function of $\Omega$.  For angular velocities below $\Omega_{crit}$ the 
terminal jet/wind speed is approximately the escape speed at the injection 
point.  When $\Omega > \Omega_{crit}$, the material is unbound and quickly 
accelerated (in a well-collimated jet, similar to MEGPL) to a speed 
such that a large fraction of the power $L_{MHD}$ is in the form of kinetic 
energy (see the right panel of Figure \ref{sim_fig}).  
It is this strong difference in the character 
of the outflows below and above $\Omega_{crit}$ that is so similar to the 
FR I/II division:  while the total power varies smoothly as $L_{MHD}$ 
exceeds $L_{crit}$, the character of the outflow varies dramatically from 
a transonic, magnetic-energy-dominated jet to a highly supersonic, 
kinetic-energy-dominated one.  

The details of the behavior of the curves in Figure \ref{sim_fig} can be 
understood qualitatively as follows.  When $\Omega < \Omega_{K}$, the 
injected plasma is no longer supported fully by rotation.  Therefore, as
soon as it emerges from the disk boundary, it dynamically contracts in 
cylindrical radius, seeking a new Keplerian equilibrium.  The jet then 
is ejected from this new, smaller, radius $R_{0} '$ with a speed similar to 
the Keplerian velocity there and a power slightly {\em larger} than that given 
by equation (\ref{nonrel_mhd_power_eq}), because of compressional increase 
in the axial magnetic field component ($B_{Z0}' \propto {R_{0} '}^{-2}$).  
If $\Omega$ is decreased further, the plasma contracts more, increasing 
the eventual jet velocity slightly.  
On the other hand, when $\Omega > \Omega_{K}$, the reverse occurs, with 
the plasma expanding to a new injection radius.  As $\Omega$ increases, 
the value of $\Omega_{crit}$ calculated from equation (\ref{omega_crit_eq}) 
changes as well, chiefly because only the radial component of the magnetic 
field is important in the expanded state ($B_{R0}' \propto {R_{0} '}^{-1}$).  
This increases 
the critical value to ${\Omega_{crit} '} = {\Omega_{K} '} {V_{esc} '}^{2} 
/ {V_{AR} '}^{2}$ (where $V_{AR} '$ is the {\em radial} component of the 
Alfv\'{e}n velocity at 
the new injection radius), which is greater than the value in equation 
(\ref{omega_crit_eq}). Finally, well above the magnetic switch (when 
$\Omega >> \Omega_{crit}$) the well-collimated jet flow gives way to a 
less-collimated radial wind, and the increase in outflow speed with 
rotation speed begins to break down.  It is not clear whether such a 
decrease at high $\Omega$ will occur in the relativistic case, where 
light cylinder effects will be important.  Relativistic MHD simulations 
are needed to investigate this behavior.

Some brief comments should be made about boundary conditions.  In these 
simulations, the kinematic and magnetic properties of the inflowing matter 
at the equator (the ``disk corona'') are fixed throughout the calculation.  
However, some authors 
allow these parameters to be determined by the solution, particularly the 
azimuthal and radial components of the field at the equator and the inflow 
velocity there (see, {\it e.g.}, \cite{ust99}).  Both of these sets of conditions 
(fixed or self-adjusting) are easy to apply numerically, and each is appropriate 
under different circumstances.  The latter is useful for generating steady state 
solutions to compare with \cite{bp82}'s self-similar, semi-analytic models.  
In this case, conditions at the base of the outflow re-adjust themselves, subject 
only to the strength of the vertical magnetic field, the density of inflow, and 
the speed and shear of the disk rotation.  The system finally reaches a steady 
state with parameters lying on (or near) the Blandford-Payne solution surface 
in $(\kappa, \lambda, \xi_{0}')$-space (see BP), with the flow varying 
smoothly from the outer regions all the way back to the equatorial boundary.  

On the other hand, fixing the inflow conditions on the equator, as we have done 
here, is perhaps more appropriate for studying the astrophysical situation where 
the magnetic field is anchored in an accretion disk (or other type of rotator). 
In that case the local magnetic field is determined by processes internal to 
the rotator, not by the outflow that originates above it.  When the boundary 
conditions are fixed, a steady flow still develops above the disk, and the 
azimuthal and radial fields and flow velocity in the corona also re-adjust.  
However, because that resulting outflow does not necessarily match onto the fixed 
disk boundary conditions, there is a transition region between the disk and the 
base of the outflow region, in which the magnetic field and velocity dramatically change in 
magnitude and direction.  It is not yet fully clear whether this transition region 
merely provides a way to connect an outflow solution that would have occurred 
anyway to the disk conditions, or whether it plays an active role in determining 
what outflow solution is chosen.  What is clear is that the values of the parameters 
chosen on the disk {\em below} the transition region {\em do} matter:  the flow changes 
dramatically with the strength of the poloidal {\em and} azimuthal field specified 
and, to a lesser extent, with the angle of the poloidal field (\cite{mpl96}; MEGPL).  
Furthermore, jet simulations in which a substantial {\em equatorial} field 
(azimuthal plus radial) is specified seem to display more collimation, 
presumably because of the strong azimuthal field that is initially present 
or that develops after a few turns of the disk.

\subsection{The Rotating Black Hole Model of Jet Production}

\label{rot_bh_subsection}

The proposed model of jet production from the accreting, rotating black 
hole system is essentially a marriage between the disk acceleration 
model of \cite{bp82} and the rotating black hole acceleration model of 
\cite{bz77} and \cite{mt82}.  
The former authors considered the case when the magnetic field threads only 
the Keplerian portion of the disk, while the latter examined the situation when it 
threads both hole and disk.  In this paper we consider the intermediate 
case when the magnetic field threads both a region of the system outside the 
static limit that does not experience appreciable frame dragging 
and a region closer to the black hole that does --- the ergosphere.
The main features of this model are the following:  (1) Unlike the BZ model, 
the magnetic field does not have to thread the horizon itself in order to 
extract the black hole rotational energy, which avoids the 
problems discussed by \cite{pc90a}, and the primary source of material in 
the jet can be a corona created by the accretion disk itself, not 
necessarily particles created in spark gaps near the horizon.  
(2) It is unlike the BP model because the disk and field 
there are frame-dragged by the hole relative to an observer at infinity, 
and thus the jet production couples well to the black hole spin, extracting rotational 
energy even though it is disk rotation that accelerates the jet.  

This model is very similar to that of PC, who studied steady state 
winds from the ergospheric region of a rotating hole.  
We note here an additional feature of such a system that occurs when time 
dependence is taken into account:  as shown in the Appendix, when a plasma 
with an appreciable poloidal magnetic field is present in the ergospheric 
region, the differential dragging of frames will increase the azimuthal 
magnetic field at the expense of the hole's rotational energy.  If, as a result 
of dynamo action, this also is accompanied by a corresponding increase in 
the poloidal component, this will have the 
effect of increasing the jet power produced near a rotating black hole to 
a value substantially larger than that for a disk around a Schwarzschild hole.

\subsubsection{MHD output power}

When a black hole is spinning rapidly, the accretion disk can extend well into 
the ergosphere --- the region where local observers must rotate with the hole 
--- with equatorial radius $R_{e} = 2 GM/c^2$.  For a nearly-maximal Kerr hole 
($j \rightarrow 1$), the last stable orbit reaches into this region --- very close to the 
horizon at $R_{H} = (1 + \sqrt{1-j^2}) \, GM/c^2$, 
where $j \equiv J/(GM^2/c)$ is the normalized angular momentum 
of the hole (equal to ``$a/M$'' in geometric units). 
Now, consider a loop of coronal magnetic field $B_{p0}$ with one foot anchored 
in the ergosphere and one outside.\footnote{This distinction is somewhat 
arbitrary, as frame dragging takes place outside the ergosphere as well, 
although at an increasingly smaller rate as one moves away from the hole, and 
it also increases within the ergosphere as one moves toward the horizon.  
The important point to note is that a field loop with its feet at different 
disk radii will receive a twist similar to that in a normal Keplerian disk, 
but driven by the differential dragging of frames in addition to any 
shearing fluid flow within the ergosphere.}  Recently \cite{hay96} and 
\cite{rom98} have shown that even in a Keplerian flow, differential rotation in 
the inner disk can modify closed coronal loops, stretching and twisting them into 
long, virtually-open vertical helical structures of the type needed to accelerate 
MHD winds and jets.  If the differential rotation is further augmented by 
frame dragging, then the work done on the field and outflow derives its power 
not only from the Keplerian rotation of the disk, but also from the rotation of 
the ergosphere itself relative to the point where the outer foot is anchored
(see the Appendix).
The system then appears to an observer at infinity as a simple magnetic 
rotator with size $R_{0} = R_{e}$, magnetic field $B_{p0}$, and angular 
velocity $\Omega \ge \Omega_{e} \approx 0.4 \Omega_{H} \approx 0.4 j/(2GM/c^3)$, 
where $\Omega_{e}$ and $\Omega_{H}$ are the angular velocity of the ergosphere 
equator and hole, respectively.  
The form of equation (\ref{nonrel_mhd_power_eq}) for a relativistic 
rotator\footnote{Equation (\ref{rel_mhd_power_eq}) also can be derived 
from equation (\ref{sigma_eq}) and the relations $\gamma_{j} = \sigma_{0}$ and 
$L_{MHD} = L_{j} \approx \gamma_{j} \dot{M}_{j} c^2$, where $\dot{M}_{j} =
4 \pi \, R_{0}^2 \, \rho_{c0} \, V_{i0}$ is the mass lost in the jet.} 
yields an MHD power for the accreting system of 
\begin{eqnarray}
\label{rel_mhd_power_eq}
L_{MHD} & = & B_{p0}^2 \, R_{0}^4 \, \Omega^2 \, / \, 4 c
\\
\label{bh_mhd_power_eq}
& = & 1.1 \times 10^{48} \, 
{\rm erg \, s^{-1}} \, \left( \frac{B_{p0}^2}{10^5 {\rm G}} \right)^2 \, 
m_{9}^2 \, j^2
\end{eqnarray}
which is a comparable power to that of the BZ mechanism for 
the same poloidal field strength.

\subsubsection{Disk magnetic field strength}

\cite{lop99} have argued that the MHD power from the accretion disk (the BP 
process,  given by equation [\ref{nonrel_mhd_power_eq}]) and that from the 
black hole (the BZ process, given by equation[\ref{rel_mhd_power_eq}] with 
$\Omega = \Omega_{H}$) are, at best, comparable, and, at worst, entirely 
dominated by the former if the hole spin is small.  They, therefore, conclude 
that the BZ process will never be important and that the spin of the black hole 
is probably irrelevant for the expected electromagnetic power of the system.  
This author agrees with the basic tenets of \cite{lop99} that the disk and 
hole power will be comparable, but not with their ultimate conclusion on the 
unimportance of the spin.  The key difference is that, in the ergosphere, 
the MHD power {\em of the disk itself} (equation \ref{bh_mhd_power_eq}) should 
depend on the spin of the black hole.  
In addition, when the hole is rotating, a much larger poloidal magnetic 
field strength is allowed than estimated by these authors, yielding black 
hole rotation-driven powers comparable to those observed in extragalactic 
radio sources.

First, consider the non-rotating Schwarzschild black hole case.  The disk will 
be a standard Keplerian one, with internal disk field given by (\cite{ss73})
\begin{equation}
\label{ss_eq}
\frac{B_{d0}^2}{8 \pi} \; \approx \; \alpha \, p_{d0}
\end{equation}
where $p_{d0}$ is the local internal disk pressure.  Now the source of a coronal 
loop of field $B_{p0}$ will be this (largely azimuthal) disk field, first 
protruding into the corona and then growing as the loop is sheared and expanded 
upward.  However, in a thin Keplerian disk, if the 
poloidal field is allowed to grow to the same order as $B_{d0}$, then the expected 
MHD power from equation (\ref{nonrel_mhd_power_eq}) would be orders of magnitude 
greater than the accretion luminosity
\begin{eqnarray}
\frac{L_{MHD}(B_{p0}=B_{d0})}{L_{acc}} & = & \frac{\alpha \, 
c_{s0}}{|V_{R0}|} 
\nonumber
\\
& = & 3.8 \times 10^{3} \, (\alpha_{\sm 2} m_{9})^{1/10} \, \dot{m}_{\sm 3}^{-1/5}
~~~~~~
\end{eqnarray}
where $c_{s0}$ is the sound speed in the disk, $V_{R0}$ is the inward accretion 
drift speed, and the notation $\alpha_{\sm 2}$ represents $\alpha / 10^{-2}$.  
(As discussed earlier, we have used a gas-pressure, 
electron-scattering-dominated disk, but the results are similar for other 
types of disks.)  There is not enough accretion luminosity to support this MHD 
power by factors of several thousand.  
As $B_{p0}$ grows to a even a small fraction of $B_{d0}$, the MHD disk power 
becomes a substantial fraction of $L_{acc}$.  At that point, a significant 
amount of angular momentum is removed from the disk by the outflow, altering 
the disk structure and enhancing the local accretion rate.  In order that the 
accretion draining the inner disk remain in a steady state with that feeding 
it from the outer disk (and not, for example, launch into a series of bursts 
that each destroy the inner disk entirely), $B_{p0}$ must be self-limited to 
about the value postulated by \cite{lop99}, or 
$B_{p0} \, \approx \, B_{d0} \, (H_{0}/R_{0}) << B_{d0}$, where $H_{0}$ is 
the half-thickness of the disk at $R_{0}$.  

However, when the central object is a rotating black hole, the situation will 
be quite different.  As before, when $B_{p0}$ grows to a fraction of $B_{d0}$, 
a significant MHD outflow will develop, exerting a braking torque on the inner 
portion of the disk.  Because of the differential frame dragging, however, even if 
{\em all} the Keplerian angular momentum were removed by the outflowing MHD 
power, the disk material still will be sheared {\em by the metric itself}, 
continuing to increase the local magnetic field strength and continuing to 
produce MHD power.  The Appendix shows quantitatively how this occurs in the 
Kerr metric, producing field enhancement everywhere in the ergospheric region.  
Tension in this growing, working field then produces a back-reaction on the 
disk plasma falling toward the hole, accelerating it relative to the rotating 
frame in a direction {\em retrograde} to the black hole spin.  By the time 
the magnetized accretion flow enters the black hole, it could have a significant 
{\em negative} angular momentum which, when added to the hole, would 
decrease the spin.  Therefore, because of frame dragging, power flows 
from hole to disk to field and finally to the outflow, exerting a braking 
torque on the hole itself.  No longer self-regulated by the accretion flow, 
$B_{p0}$ in the ergospheric region above the disk now can continue to grow 
at the expense of the hole rotational energy to order $B_{d0}$, 
given by the disk solutions with a Kerr metric (\cite{nt73}).  For the gas 
pressure/electron-scattering accretion model assumed here, this value is
\begin{equation}
B_{p0} \; \approx \; B_{d0} \; = \; 3.4 \times 10^{4} \, {\rm G} \, 
\alpha_{\sm 2}^{1/20} \, m_{9}^{-9/20} \, \dot{m}_{\sm 3}^{2/5}
\end{equation}
Equation (\ref{bh_mhd_power_eq}) then becomes 
\begin{equation}
\label{lmhd_eq}
L_{MHD} \; = \; 1.3 \times 10^{47} \, {\rm erg \, s^{-1}} \, 
\alpha_{\sm 2}^{1/10} \, m_{9}^{11/10} \, \dot{m}_{\sm 3}^{4/5} \, j^2
\end{equation}
(This differs from equation (16) of \cite{ms96} by a factor of $\alpha$, 
from equation (\ref{ss_eq}), and a factor of $\sim 0.4^{2}$ from the 
slightly smaller ergospheric rotation rate.)  
While $L_{MHD}$ appears to scale with $m_{9}$ and $\dot{m}$ in a 
manner similar to an accretion luminosity, it is important to note 
that it does so only because of the scaling on $B_{d0}$.  The 
energy liberated is derived mainly from the rotation of the 
hole itself and potentially much larger than the accretion luminosity 
($L_{acc} \approx 10^{43} {\rm erg s^{-1}} \, m_{9} \, \dot{m}_{\sm 3}$).

\subsubsection{Observed radio power}

To turn $L_{MHD}$ into an observed radio power, we note that, over long 
periods of time, the central engine may generate a jet only a fraction $\zeta$ 
of the time.  Furthermore, only a fraction $\epsilon$ of this average power 
is radiated at GHz frequencies in a 1 GHz bandwidth and observed with standard 
radio telescopes.  
From millimeter and optical observations of repeated outbursts of radio 
galaxies and quasars (see, {\it e.g.}, \cite{rob92}), we estimate typical 
duty cycle values to be $\zeta \sim 0.01 - 0.1$. 
The radio efficiency is also very uncertain, but we note that for 
the microquasars GRO J1655-40 and GRS 1915+105, estimates of $\epsilon$ 
are of order $10^{-5}$ or so, just for the instantaneous moving jet 
(\cite{m96}).  A canonical value of $\epsilon \sim 10^{-3}$ was chosen for 
these calculations by comparing equation (\ref{lmhd_eq}) with the maximum 
observed radio source power ($10^{27} \, {\rm W \, Hz^{-1}}$) 
for a radio galaxy that should have a $10^{9} \, M_{\sun}$ 
hole in its nucleus ($M_B \approx -21.0$).  This value for $\epsilon$ is 
consistent with total conversion rates of $\sim 0.01-0.1$ (see, {\it e.g.}, 
\cite{leahy91}), with only a fraction of that radiated in a 1 GHz bandwidth.  
Scaling to these values, 
the calculated radio power for different black hole spins is 
\begin{eqnarray}
P_{rad} & = & \epsilon \, \zeta \, L_{MHD} \, / \, 10^{9} \, {\rm Hz}
\\
& = & 1.3 \times 10^{27} \, {\rm W \, Hz^{-1}} \; m_{9}^{11/10} j^2 
\nonumber
\\
\label{pmhd_eq}
& & \times \; \left( \epsilon_{\sm 3} \, 
\zeta_{\sm 1} \, \alpha_{\sm 2}^{1/10} \, \dot{m}_{\sm 3}^{4/5} \right)
\end{eqnarray}
where $\dot{m}$ is still the average {\em outburst} accretion rate only, 
not that averaged over the entire duty cycle.  

\subsection{The Magnetic Switch as the FR I/II Break}

The final step in the model is to 
identify the magnetic switch as the cause of the FR I/II break.  
That is, one should observe a distinct change in the character of the jet 
produced when $L_{MHD} = L_{crit}$, or when the radio power attains the 
critical value
\begin{eqnarray}
\label{pfr_eq}
P_{rad,mod}^{FR} & = & \epsilon \, \zeta \, L_{crit} \, / \, 10^{9} \, {\rm Hz}
\end{eqnarray}
To compute an observed break power 
we need an expression for the corona density $\rho_{c0}$ 
at $R_{0}$.  Of course, coronal physics is poorly understood, even for the 
sun.  Nevertheless, we can obtain a rough estimate by assuming that the corona 
is produced locally by the accretion disk, so the disk and corona 
densities should scale together as\footnote{Another possible scaling would be with
both $\rho_{d0}$ and the internal disk temperature $T_{d0}$ ({\it i.e.}, 
with $p_{d0}$).  This yields a slightly weaker $M_{R}$ slope (-0.55) 
in equation (\ref{mod_olwl_eq}) below, but still within the observational 
errors.}
\begin{equation}
\label{rhoc_eq}
\rho_{c0} \; \approx \; \eta \, \rho_{d0}
\end{equation}
where $\rho_{d0}$ is the internal disk density at $R_{0}$ and $\eta << 1$ 
is a parameter which, while not necessarily a true constant, is assumed to 
vary sufficiently slowly with the other parameters that it does not affect 
significantly the scaling of our results with $m_{9}$, $j$, and $\dot{m}$.  
The parameter $\eta$ is expected to be quite small; in the 
solar corona, for example, $\eta_{\sun} \equiv \rho_{c\sun} /
\rho_{\sun}(\tau=1) \sim 10^{-8} - 10^{-11}$ in the region between 
the photosphere and $2 R_{\sun}$.  If the corona is a pair plasma, 
$\eta$ (which is a mass ratio) could be very small indeed.  
Combining equations (\ref{lcrit_eq}), (\ref{pfr_eq}), and (\ref{rhoc_eq}), 
and again using the gas pressure/electron-scattering disk, one obtains 
\begin{eqnarray}
\label{th_frbr_eq}
P_{rad,mod}^{FR} & = & 4.8 \times 10^{25} \, {\rm W \, Hz^{-1}} \; 
m_{9}^{13/10} 
\nonumber
\\
& & \times \; \left( \epsilon_{\sm 3} \, \zeta_{\sm 1} \, \alpha_{\sm 2}^{-7/10} \, 
\dot{m}_{\sm 3}^{2/5} \, \eta_{\sm 11} \right) 
\end{eqnarray}
This is the magnetic switch model prediction for the locus of the FR I/II break 
as a function of black hole mass.

\section{Applications and Discussion}

This section applies the model to the FR I/II break in the radio-optical plane, 
shows that the rotationally-triggered magnetic switch {\em is} consistent with 
long source morphology lifetimes, and addresses the more general questions of 
evolution and radio loud {\it vs.} radio quiet sources.  The important features 
of the model and its applications are shown in figure \ref{roplane_fig}.  

\label{appl_section}

\subsection{The ``OLWL'' Relation in the Radio-Optical Plane}

\label{olwl_subsection}

\cite{ol89}, \cite{ow91}, and \cite{lo96} have shown that FR I and FR II 
type radio galaxies occupy different regions of the $\log P_{rad,obs} - 
M_{R,24.5}$ plane, where $\log P_{rad,obs}$ is the observed radio power at 
1400 MHz (in ${\rm W \, Hz^{-1}}$) and $M_{R,24.5}$ is the galaxy optical 
red magnitude measured to a surface brightness of $24.5$ magnitudes per 
square arcsecond.  In particular, the tendency for a galaxy of a given 
magnitude to transition from an FR I to an FR II appears to be a strong 
function of optical magnitude. The dividing line is given approximately by
\begin{equation}
\label{obs_olwl_eq}
\log P_{rad,obs}^{FR} \; = \; -0.66 \, M_{R,24.5} \; + \; 10.35
\end{equation}
or $P_{rad,obs}^{FR} \propto L_{opt}^{1.65}$.  The error in the coefficient 
of $M_{R,24.5}$ is at least $\pm 0.1$ and in the intercept at least 
$\pm 0.3-0.5$.  The Owen-Laing-White-Ledlow 
(OLWL) relation is a more precise statement of the Fanaroff \& Riley class 
division, so care must be taken when classifying radio galaxies based on 
their radio power only.  There exist FR II radio galaxies with quite small 
powers (occurring in less bright elliptical galaxies), as well as FR I objects 
with very high radio power (occurring in very bright galaxies).  

From the magnetic switch model for the FR I/II break, it is possible to 
calculate a theoretical relation for the observed break power to compare with 
equation (\ref{obs_olwl_eq}).  Using equation (\ref{th_frbr_eq}) to define the 
break, equation (\ref{bh_mass_vs_gal_mag_eq}) to convert black hole mass to 
galaxy bulge blue optical magnitude, and a standard $B-R \sim 1.5$ to convert 
to elliptical galaxy blue magnitudes to red, one obtains 
\begin{eqnarray}
\log P_{rad,mod}^{FR} & = & -0.65 \, M_{R} \; + 11.1 
\nonumber
\\
\label{mod_olwl_eq}
& & + \; \log (\epsilon_{\sm 3} \, \zeta_{\sm 1} \, \alpha_{\sm 2}^{-7/10} \, 
\dot{m}_{\sm 3}^{2/5} \, \eta_{\sm 11}) ~~~~~
\end{eqnarray}
Given the uncertainties in the observed slope, in the scaling of the coronal 
density, and in the parameters $\epsilon$, $\zeta$, and, especially, $\eta$, 
the good numerical agreement between equations (\ref{obs_olwl_eq}) and 
(\ref{mod_olwl_eq}) is rather fortuitous, of course.  The important points to 
note are as follows:
\begin{itemize}
\item{The basic trend that brighter galaxies, with their more massive black 
holes, should undergo the magnetic switch at a higher jet power is a general 
feature of the model (see figure \ref{roplane_fig}).  The ultimate source of 
this trend is that the critical 
jet power is proportional to the size of the ergosphere squared times the disk 
coronal density there.  If this density were uniform for all black holes, it 
would produce a trend slightly steeper than $P_{rad}^{FR} \propto L_{opt}^2$,
because the galaxy $M/L$ ratio increases with galaxy $L_{opt}$ (\cite{faber87}); 
however, as densities tend to decrease as the size of the accreting system 
increases, the trend should be {\em less} steep than this.}
\item{Although the uncertainties are great, the intercept is consistent with 
the values of the parameters chosen here, in particular with the very low 
value of $\eta$.  That is, {\em the observations are entirely consistent with 
the jet material originating in the coronae of the black hole accretion disks}, 
and grossly {\em in}consistent --- by more than ten orders of magnitude --- 
with the jet ejecting a large fraction of the accretion disk itself.}
\end{itemize}

Incidentally, the present model is {\em not} inconsistent with observations that 
FR I jets may decelerate on kiloparsec scales (\cite{laing96}).  Indeed, it 
suggests a reason 
for such deceleration to occur more readily in FR I sources.  Jets with 
$L_{MHD} < L_{crit}$ are transonic in the simulations here and in MEGPL. 
Such jets are more prone to deceleration than highly supersonic flows 
(\cite{bick85}).  Note that deceleration of relativistic parsec-scale jets 
to nonrelativistic flow on kiloparsec scales is needed in {\em any} 
model, magnetically switched or not, to explain head-tail and wide-angle-tail 
sources in clusters, whose jets are bent by motion of only a few thousand 
${\rm km \, s^{-1}}$ through the intracluster medium.

\subsection{Black Hole Spindown and the Early Evolution of FR II Radio Sources}

Because the black hole spin is the assumed source of power, there should 
be a critical spin $j_{crit}$ --- corresponding to a critical hole 
angular velocity $\Omega_{H,crit}$ --- such that the magnetic switch is 
triggered when the hole is spun up to this value.  Equating
$P_{rad,mod}^{FR}$ and $P_{rad,mod}$ (or $L_{MHD}$ and $L_{crit}$) one 
obtains
\begin{equation}
\label{jcrit_eq}
j_{crit} \; = \; 0.20 \; \alpha_{\sm 2}^{-2/5} \, m_{9}^{1/10} \, 
\dot{m}_{\sm 3}^{-1/5} \, \eta_{\sm 11}^{1/2}
\end{equation}
Now, while this relation is a very weak function of $m_{9}$ and $\dot{m}$, 
and $\alpha$ is considered to be known to within a factor of ten or better,
the exact value of $j_{crit}$ is still very uncertain because of the
parameter $\eta$.  Indeed, a large value (dense corona) would make 
$j_{crit} > 1$, implying that even a maximal Kerr hole would not trigger 
the switch.  Below observational tests are proposed to 
confirm the low choice of $\eta$ in this paper.

Note that, independent of the magnetic switch model, if the source of 
the radio jet power is, indeed, the black hole rotational energy, then 
simple arguments show that at the FR I/II transition, $j$ must be of 
order $0.1$ in any case.  From the OLWL results and the bivariate 
radio luminosity function itself, it appears that, for a galaxy of a 
given mass, the maximum possible radio power is about 100 
times the power at the FR break.  Therefore, for any model 
where $P_{rad} \propto j^2$, we find that 
\begin{equation}
j_{obs}^{FR} \; \approx \; \left( \frac{P_{rad}^{FR}}{P_{rad}^{max}} 
\right)^{1/2} \; \sim \; 0.1
\end{equation}

The magnetically-switched, rotating black hole model further predicts that 
FR II sources contain the seeds of their own destruction.  The jet extracts 
rotational energy from the black hole until $j < j_{crit}$, 
whereupon the jet transitions to an FR I source.  Now, in the absence of 
external effects, the spindown 
rate is proportional to $j^2$, and therefore the hole rotational energy
\begin{eqnarray}
E_{rot} & = & M c^2 \; \left[ (1 \, + \, j^2/4)^{1/2} \, - \, 1 \right]
\nonumber
\\
& \approx & 1.6 \times 10^{62} \, {\rm erg} \; m_{9} \, j^2 ~~~~~ ,
\end{eqnarray}
rendering the spindown an exponential process, with a time scale
\begin{eqnarray}
\label{spindown_time_eq}
\tau_{spindown} & = & \frac{E_{rot}}{\zeta \, L_{MHD}} ~~~~~ 
\nonumber
\\
& = & 4.1 \times 10^{8} \, {\rm yr} \; \alpha_{\sm 2}^{-1/10} \, 
m_{9}^{-1/10} \, \dot{m}_{\sm 3}^{-4/5} \, \zeta_{\sm 1}^{-1} 
~~~~~~~~~~
\end{eqnarray}
where $\zeta$ is the duty cycle parameter defined earlier.  
This spindown time is $\lesssim 10^{9} {\rm yr}$ time scale 
found for the cosmic evolution of powerful extragalactic radio 
sources (\cite{schm72}; \cite{wj97}).
A plausible scenario, then, is that a fraction of black holes are born 
with very high spin (FR II sources) and subsequently experience a strong 
density evolution, disappearing after $\tau_{spindown}$ or so.  Note that 
the local FR I space density exceeds that of FR II sources at all epochs, so 
that expired FR II sources could easily be hiding in the FR I population (\cite{jw99}; 
Jackson, private communication).  

Note also that, like the accretion time
\begin{equation}
\tau_{acc} \; \equiv \; M \, / \, \dot{M} \; = \; 4.5 \times 10^{12} {\rm yr} 
\; \dot{m}_{\sm 3}^{-1}\, \zeta_{\sm 1}^{-1} 
\end{equation}
$\tau_{spindown}$ is virtually independent of black hole mass (although much 
shorter).  Therefore, in the absence of other 
effects, even {\em stellar} mass black holes like GRO J1655-40 or GRS 1915+105 
are expected to retain their spin over $\sim 10^{9} \, {\rm yr}$ as they use
their rotational energy to drive radio jets.  Such a long spin lifetime 
should have implications for models for galactic microquasars. 

During the entire lifetime of the radio source, an energy comparable 
to $E_{rot}$ should be liberated into the lobes.  That is, extended 
radio sources are the remnants of formerly rapidly-rotating black holes.
This amount of energy is consistent with, and in fact somewhat 
larger than, the total energy content of extended radio sources in
relativistic particles and magnetic fields (\cite{dey76}).  This allows for 
a significant fraction of this energy to have been converted to other 
forms ({\it e.g.}, heating of the intergalactic medium) or radiated away.

\subsection{Little Evolution for FR I (and FR II) Sources at Late Times?}

The fact that $E_{rot}$ is proportional to $j^2$, and not to just $j$, 
for example, means that the spindown time is the same no matter how 
rapidly or slowly the hole is rotating.  The model predicts therefore, 
that after turning into an FR I source, the object will 
continue to display pure luminosity evolution on a $\tau_{spindown}$ time 
scale.  However, recent analyses of the radio source counts and redshift 
data (\cite{up95}; \cite{wj97}) conclude that the FR I population evolves 
little or not at all.  Modeling this sort of behavior is very difficult, 
especially in a universe where everything involved with AGN seems to evolve 
on at least a cosmological time scale (galaxy formation and mergers; the 
rate of stellar death [which may feed gas into the nucleus]; even radio 
quiet quasars; {\it etc.}).  Below are some ideas for halting, or at least 
slowing, the evolution of FR I sources in this model; however, much more 
work needs to be done on this problem:
\begin{itemize}
\item {{\em Constant low-powered jet from outside the ergosphere.}  
After spinning down, the black hole loses most of its ergosphere and 
its ability to create a large poloidal field strength in the disk.  
The field strength is expected to drop to the value suggested by 
\cite{lop99}, yielding an expected jet power of 
\begin{equation}
\label{min_mhd_power_eq}
L_{MHD}^{min} \; = \; 9 \times 10^{40} \, {\rm erg \, s^{-1}} \; 
\alpha_{\sm 2}^{-1/10} \, m_{9}^{9/10} \, \dot{m}_{\sm 3}^{6/5} 
\end{equation}
which, with efficiency factors similar to those used above, could 
generate radio emission of up to $\sim 10^{21} \, {\rm W \, Hz^{-1}}$. 
A radio source of this power is not expected to evolve much, unless 
$\dot{m}$ evolves.  However, this is a very weak power --- much smaller 
than many FR I radio galaxies, which also presumably do not evolve much.
Furthermore, if the accretion events have rather random angular momenta, 
then this minimum jet has no preferred direction.  It is likely to 
point in random directions with each event, never developing into an 
extended source, and perhaps appearing radio quiet.}
\item {{\em Periodic spinup of the black hole.}  On the other hand, if 
the short-term accretion events have systematically the same angular 
momentum direction, as might be the case for a galaxy collision or merger, then 
while they would keep the above weak jet oriented in the same direction, 
they also would tend to re-orient and spin up the hole (\cite{wc95}; 
\cite{np98}).  
This, or the direct merger of two black holes, would re-kindle the 
ergospheric jet.  Then, if the galaxy merger rate evolved on a very 
long time scale (say, $\gtrsim 5 \times 10^{9} \, {\rm yr}$), the 
shorter time scale spindown process would become slave to the spinup 
process, with the average radio power being a function of the rate 
at which angular momentum is accreted onto the hole.  At low redshifts, 
then, both FR I and the occasional re-kindled FR II objects would 
evolve on the longer merger evolution time, not $\tau_{spindown}$.}
\end{itemize}

\subsection{Radio Loud {\it vs.} Radio Quiet?}

The model proposed here deals mainly with the difference between two 
radio loud populations --- the FR I and II sources.  However, it also 
has implications for the difference between radio loud and quiet 
quasars as well.  In particular, it suggests that radio quiet 
objects should be identified with very slowly-rotating, nearly-Schwarzschild 
black holes.  At first, the minimum jet power (equation 
\ref{min_mhd_power_eq}) appears to be a problem with this suggestion, 
as it predicts that all active galaxies should produce some sort of 
extended radio emission.\footnote{Of course, as discussed earlier, 
due to a lack of a consistent long-term jet direction, this power may 
never manifest itself as an extended radio source.  Here it will 
be assumed that it does; if not, the conclusions will be stronger.}  
However, even if an AGN does produce an extended radio source power of 
\begin{eqnarray}
\label{pmin_eq}
P_{rad}^{min} & \leq & 9 \times 10^{20} \, {\rm W \, Hz^{-1}} \;
m_{9}^{9/10} 
\nonumber
\\
& & \times \; \left ( \epsilon_{\sm 3} \, 
\zeta_{\sm 1} \, \alpha_{\sm 2}^{-1/10} \, \dot{m}_{\sm 3}^{6/5} \right)
\end{eqnarray}
the apparent flux would be quite small:  less than a few ${\rm \mu Jy}$ 
for Seyfert galaxies at $z > 0.03$ and quasi-stellar objects at 
$z > 1$, and a few millijanskies total for nearby spiral galaxies.  
This amount of extended emission in spirals could be hidden easily by 
emission from supernova remnants, {\it etc.}.  Therefore, there should 
be a minimum spin rate $j_{min}$ for the hole such that when 
$j < j_{min}$, it will appear radio quiet. 
While detailed simulations of jet production from rotating 
black holes, plus observational selection factors, need to 
be taken into account to determine a precise value of $j_{min}$, we 
can estimate it here by equating the MHD power generated by the ergospheric 
and Keplerian disk jets (equations \ref{lmhd_eq} and 
\ref{min_mhd_power_eq}) and solving:
\begin{equation}
\label{jmin_eq}
j_{min} \; = \; 0.83 \times 10^{-3} \; \alpha_{\sm 2}^{-1/10} \, m_{9}^{-1/10} \, 
\dot{m}_{\sm 3}^{1/5} 
\end{equation}
For spins slower than this, the wind/jet from the Keplerian portion of the 
disk outside the ergosphere will dominate the output MHD power.

Even if the minimum jet power turns out not to be a problem, one still 
needs to explain --- quantitatively as well as qualitatively --- why the 
radio power distribution is bimodal:  radio quiet objects appear to be 
a separate class, not merely the fainter version of the powerful radio 
sources (\cite{kell89}).  One possible solution to this problem is the 
re-kindling the radio source through galaxy mergers or collisions.  
This is not a new idea (\cite{wc95}), but the model presented here, especially 
equation (\ref{spindown_time_eq}), gives a physical and quantitative 
basis for it.  In the early universe, holes would have been born 
with a spectrum of rotational speeds.  Those rotating rapidly would 
have produced extended radio sources, including holes in some 
(pre-)spiral galactic bulges, while those rotating slowly 
would have been radio quiet.  All of these objects should have spun 
down in a few $\tau_{spindown}$ times and become radio quiet, according 
to the model.  However, a subgroup --- those in galaxies that 
experience repeated collisions and mergers --- will be spun up again, 
at least partially to FR I class and very occasionally to FR II strength, 
re-kindling the radio source.  This scenario is 
consistent with powerful radio sources appearing predominately in 
elliptical galaxies, which now are believed to form from, 
or be strongly influenced by, mergers. 
An important corollary is that, early in their history, even the bulges of 
a significant fraction of spiral galaxies should have undergone a one-time 
extended radio source phase, lasting a few $\tau_{spindown}$ times and 
producing both FR I and II morphologies, but with considerably weaker 
luminosities than those associated with present-day giant elliptical radio 
sources.  

The radio loud/quiet 
problem thus may be one of environment rather than being intrinsic 
to the system.  However, before adopting such a simple model, it will 
be important to understand the strange case of the microquasar 
GRO J1655-40, which can be either radio loud or radio quiet during its 
X-ray outbursts (\cite{tav96}).  It is not clear yet whether this is simply 
a short-term phenomenon or is fundamentally related to the {\em macro}quasar 
radio loud/quiet problem.  So far, no model, including the one in this paper, 
appears able to explain completely this object's behavior.

\subsection{Secondary Evolution in Accretion Rate and the Lack of FR I 
Quasars}

It has been assumed in this paper that on a spindown time the average 
$\dot{m}$ is fairly independent of black hole mass and remains at 
roughly $\sim 10^{-3}$.  However, on a cosmological time scale this is not 
likely to be the case.  Indeed, there is evidence (\cite{wj97}; \cite{jw99}) that 
some sources (class `A') have a high accretion rate, indicated by strong, 
broad line emission, while others (class `B') have a low (perhaps ADAF-like) rate.
For the case where the latter objects are core-dominated and 
appearing as BL Lac objects, Wall \& Jackson note that at least 
some must have FR II sources as their parent population, not just FR I sources.
One possible explanation of this is that the average $\dot{m}$ may have been 
much higher in the early universe when the merger rate was high, 
and lower, perhaps even ADAF-like, now.  Indeed, it appears that it 
is only the class `A' objects that display rapid evolution (\cite{jw99}).
Then, present-day re-kindled FR I 
and II sources would both have a low average accretion rate, as well as a low 
evolution rate, and both have 
the potential for appearing as a BL Lac object if viewed end-on.
Not all FR II sources seen this way would have to appear as quasars. 

Ascribing a high accretion rate to class `A' objects solves 
another puzzle concerning quasars --- the fact that few, if any, radio loud 
quasars display FR I morphology.  For $\dot{m} \sim 1$, equation 
(\ref{jcrit_eq}) gives $j_{crit} \approx 0.05$, implying that even 
slowly-rotating black holes would produce an FR II morphology when 
the accretion rate is high.  The right panel of figure \ref{roplane_fig} 
illustrates that the FR I portion of the quasar diagram is significantly 
smaller than that for radio galaxies;  the exact size of the region depends 
critically on the definition of radio quietness.

\section{Tests of the Model}

\label{tests_section} 

\subsection{Observational Tests}

This paper has shown that the magnetic switch mechanism, triggered 
by a rapidly spinning black hole, is consistent with a variety of 
observations concerning the FR I/II break.  However, Bicknell (1995) 
has shown that an ISM model is consistent as well.  In order 
to distinguish between these two, the following observational 
tests are proposed.  The current model predicts the following:
\begin{enumerate}
\item {The ensemble average of VLBI jet speeds for sources with FR I 
{\em morphology} should be distinctly slower than that for FR II 
sources.  Until detailed general relativistic MHD simulations of very 
tenuous coronae in Kerr black hole ergospheres can be done ({\it cf.} 
\cite{koide98b}), the exact 
values of the two mean velocities will be uncertain.  The FR II mean 
should be highly relativistic ({\it i.e.}, $\gamma \sim 5-10$ or greater; 
see MEGPL).
However, the FR I value may be more than mildly relativistic itself, 
as the escape velocity from the ergosphere is very close to $c$.
Note the emphasis on FR morphology, rather than radio power.  Radio 
power should {\em not} be used to distinguish between FR class in this 
test, unless it can be shown (using the OLWL relation) that the galaxy 
in question lies well within either the FR I or FR II region of the 
radio-optical diagram.  Faint galaxies can have low-power 
radio sources that still have FR II morphology, and vice-versa.}
\item {Direct measurement of the accretion disk corona density in the 
ergosphere, using X-ray observations for example, should yield rather 
low mass densities ($\sim 10^{-15} {\rm g \, cm^{-3}}$), perhaps 
either a $10^{12} \, {\rm cm^{-3}}$ $e^{+}e^{-}$ pair plasma or a 
very tenuous $10^{9} \, {\rm cm^{-3}}$ $p^{+}e^{-}$ corona. 
These low values are needed for $L_{crit}$ to be such that a radio 
power of $P_{rad,mod}^{FR} \sim 10^{25} {\rm W Hz^{-1}}$ is generated at the FR break.  
Significantly higher measured densities would render the corona too 
heavy for the magnetic switch to be triggered by any reasonable jet 
power, and, in this model, always would predict slow jet velocities.}
\item {Recently, methods have been developed to estimate the spin of 
black holes by measuring the rotation rate of the innermost part of 
their accretion disks (\cite{zhang97}; \cite{sobc98}).  If such 
methods can be applied to 
extragalactic radio sources, one should find that the ensemble 
average of black hole spins for sources with FR I morphology should be 
distinctly lower than for those with FR II morphology.}
\item {There should exist a population of high-redshift, extended 
radio galaxies associated with the bulges of spiral or pre-spiral galaxies.  
They should include both FR I and II morphologies, but be weaker copies 
of the giant sources associated with ellipticals.  
The radio sources themselves should be readily detectable (for $m_{9} = 10^{-2}$ 
and a redshift of 3, $P_{rad} \sim 10^{22-25} {\rm W \, Hz^{-1}}$ or 
$S \sim 4-4000 {\rm \mu Jy}$), but their optical counterparts would be very 
faint if undergoing an initial burst of star formation ($M_{R} \sim 25$ for a 
$10^{10} M_{\sun}$ bulge at a similar redshift; \cite{m76}) and even fainter 
($M_{R} \sim 30$) if not. Preliminary estimates of their numbers are on 
the order of a few hundred to a few thousand per square degree --- considerably 
smaller than the larger number of starburst galaxies at faint fluxes (\cite{wind95}).}
\end{enumerate}

The first and third of these tests would provide support for the magnetic 
switch mechanism for the FR I/II transition.  The second would support the 
choice for a small $\eta$.  The third and fourth provide evidence for the 
proposal that the black hole spin is the switch triggering mechanism and the 
ultimate source of the radio power.

\subsection{Numerical Tests}

A number of numerical tests will be almost as important as the observational 
ones:
\begin{enumerate}
\item {Firstly, as suggested by MEGPL, the magnetic switch needs to be 
confirmed by other groups performing numerical simulations, especially with 
a general relativistic MHD code, even if just for Schwarzschild black holes.}
\item {Secondly, general relativistic MHD simulations in Kerr geometry are 
needed to confirm that a substantial jet, whose power is directly related 
to the black hole spin, can be produced simply by threading the ergosphere 
with a magnetic field that is anchored further out.  A simple vertical 
field should be sufficient initially, but simulations similar to those of 
\cite{rom98}, with closed coronal loops, would provide a more realistic 
test.}
\item {Finally, it will be important to reproduce in Kerr geometry the 
simple simulations performed in this paper, showing that increasing the 
black hole spin eventually triggers the magnetic switch.  It also will 
be important to determine the speed, Mach number, and collimation 
properties of the outflow when the black hole spin is well above 
$\Omega_{crit}$ to see if the flow de-collimates (as it did in these 
non-relativistic simulations) or remains well-collimated.} 
\end{enumerate}

\section{Conclusions}

The magnetically-switched, rotating black hole model for the FR I/II break 
is consistent with current black hole accretion models for active galactic 
nuclei, with MHD simulations of jet production, and with relativistic wind 
theory.  It predicts that a sharp transition in jet morphology should occur 
with increasing radio power, as observed, and that the high luminosity jets 
should be fast and kinetic energy dominated.  It yields the correct trend 
of this transition with galaxy luminosity (larger galaxies should transition 
at larger radio power), and is consistent with the actual value of the 
FR I/II break radio power, {\em if} the average density of accretion disk 
coronae in black hole ergospheres is rather low ($\sim 10^{-15} \, {\rm g \, cm^{-3}}$).  
The model suggests a long-term memory mechanism --- the black hole spin --- 
for assuring that radio sources maintain the same morphology over periods of 
time that are at least as long as the flow time from galactic center to lobe 
($10^{6-7} \, {\rm yr}$).  In that regard, it builds on an earlier suggestion 
(\cite{rees78}) that black hole spin is the mechanism determining the 
{\em direction} of the radio jet (although the black hole spin itself may ultimately 
be tied to the angular momentum of mergers or collisions; see \cite{np98}).  
Because black hole rotational energy drives 
the radio jet, the model predicts spindown and, consequently, cosmic evolution 
of the FR II space density on a time scale of order $10^{9} \, {\rm yr}$ or less. 
It also suggests a possible model for radio quiet quasars (Schwarzschild black 
holes) and for why radio loud quasars display predominantly FR II morphology. 
Finally, the total amount of rotational energy liberated by 
the jets from a rapidly-rotating hole is more than adequate to account for the 
entire energy content of a typical extended radio source.

Several observational and numerical experiments have been proposed to test this 
model.  The most important of these is the VLBI jet velocity test, in which 
objects with FR I morphology should have a distinctly slower ensemble-averaged 
jet speed than that for FR II sources.  Other tests involve a similar effect 
for the black hole spin, a prediction for the typical density of a black 
hole accretion disk corona, and a search at high redshift for a population of 
faint extended radio sources associated with spiral or pre-spiral bulges.  
Proposed numerical experiments include general 
relativistic simulations confirming that the magnetic switch process works 
and that output MHD power depends critically on black hole spin.

{\bf Note added in proof:}  In discussing the literature on the expected power 
output of a rotating black hole, the author omitted an important reference.  
P. Ghosh \& M. A. Abramowicz (1997, {\it Mon. Not. Royal 
Astron. Soc.} {\bf 292}, 887) also discussed this subject at length, assuming, as is 
done here, that $B_{p0} \approx B_{d0}$, and obtained expressions for the MHD power 
output of a rotating hole similar to those in the present paper.  (Note that these 
authors took all of this power to be due to the Blandford-Znajek process occurring 
very near the horizon.)  However, the absolute magnitude of their power, is about 
two orders of magnitude smaller than that in equation (12). Part of this 
difference is due to a factor of $8$ in the definition of the MHD power (which was 
discussed by these authors), but most of it is due to the present paper recognizing 
that the accretion disk can extend well interior to $6GM/c^2$ and even inside the 
ergosphere for rapidly rotating black holes.  This yields much larger 
magnetic field strengths near the static limit and black hole horizon.  
In fact, one way in which Ghosh \& Abramowicz (1997), Livio {\it et al.} (1999), and this paper 
can be brought into rough agreement is to 1) extend the disk into the ergosphere (which 
would increase the power computed by Ghosh \& Abramowicz) and 2) treat that disk as 
an advection-dominated one with $H_{0}/R_{0} \approx 1$, yielding $B_{p0} \approx B_{d0}$ 
(which would increase the power calculated by Livio, Ogilvie, \& Pringle).  

\acknowledgments

The author is pleased to acknowledge the following:  discussions with D. Jones, 
S. Koide, and D. Murphy;  comments made by R. D. Blandford and E. S. Phinney during a 
seminar on this subject by the author; and helpful e-mail exchanges with E. Fomalont, 
C. Jackson, and R. Khanna.  
This research was carried out at the Jet Propulsion Laboratory, California 
Institute of Technology, under contract to the National Aeronautics and Space 
Administration.

\begin{appendix}{}

\label{msd_appendix}

\section{Metric-Shear-Driven MHD Dynamos}

This appendix shows that shear in a general relativistic metric, like that displayed 
in Kerr geometry, can serve to enhance the local magnetic field strength.  
Magnetized plasma accreting onto a rotating black hole, therefore, will extract energy 
from the hole, converting it into enhanced local magnetic field strength.  
The effect occurs even in absence of shear in the plasma fluid flow itself and even 
if the magnetic field lines do not thread the black hole.

In a classical MHD flow the magnetic field evolution equation, derived from 
Faraday's and Ohm's laws with infinite conductivity, is
\begin{equation}
\label{nonrel_b_evol_eq}
\frac{\partial {\bm B}}{\partial t} \, + \, {\bm v}{\bm \cdot}{\bm \nabla}{\bm B} 
\; = \; {\bm B}{\bm \cdot}{\bm \nabla}{\bm v} \; - \; 
{\bm B}\, {\bm \nabla}{\bm \cdot}{\bm v}
\end{equation}
where ${\bm B}$ and ${\bm v}$ are the magnetic field and velocity vectors.  
For an initially poloidal field and rotational flow only, the $\phi$ 
component of this equation becomes, in spherical-polar coordinates, 
\begin{equation}
\label{nonrel_bp_evol_eq}
\frac{\partial B_{\phi}}{\partial t} \; = r \, sin \theta \, \left[
B_{r} \, \frac{\partial \Omega}{\partial r} \; + \; 
B_{\theta} \, \frac{1}{r} \, \frac{\partial \Omega}{\partial \theta}
\right]
\end{equation}
where $\Omega \equiv v_{\phi}/(r \, sin \theta)$ is the angular velocity field 
of the flow.  Thus, if there is shear in 
a classical rotating magnetized fluid flow, a poloidal magnetic field will 
serve as a seed for generation of a potentially large azimuthal magnetic field 
that grows with time.  This ``omega-effect'' is a necessary (though not 
sufficient) condition for an ``alpha-omega'' dynamo to exist (\cite{r93}).
According to Cowling's theorem (\cite{c34}), for a full dynamo, there also must 
be non-axisymmetric poloidal fluid motions (due 
to turbulence, for example) that enhance the poloidal magnetic field as well.  
Such turbulent motions are thought to be a common occurrence when the rotating 
shear flow occurs in a gravitational field (\cite{stone96}; \cite{bnst96}).  
So, the presence of an omega-effect in a gravitational field is a strong 
indication that an alpha-omega dynamo may be operating and generating a 
substantial magnetosphere.  

The theory of gravito-magnetic dynamos in the Kerr metric has been developed by 
\cite{kc96a}, raising the possibility that even an {\em axisymmetric} flow can 
have growing, self-excited modes.  Although such modes so far have not been found 
in kinematic numerical simulations (\cite{brand96}; \cite{kc96b}), the 
non-validity of Cowling's theorem in a general relativistic metric has been 
analytically confirmed by \cite{n97}, who showed that there are growing 
axisymmetric modes when steep gradients exist in the plasma angular velocity 
$\Omega$ relative to the rotating space.  Here, however, we consider the simpler 
situation where the plasma flow itself does not shear, but there still is shear 
due to the general relativistic effect of frame-dragging by a rotating black hole.  
While this does not demonstrate the existence of a dynamo in the system, it does 
satisfy several of the necessary conditions for such --- in particular, an 
omega-effect occurring in a gravitational field.  

To compute the evolution of the azimuthal magnetic field, 
we use two different approaches --- one that uses the 
covariant four-dimensional formalism of \cite{mtw73} (MTW)  and one that uses the 
``3+1'' decomposition of \cite{tpm86} (TPM).  Both give insight into the magnetic 
field enhancement process.  
The general relativistic analogy of equation (\ref{nonrel_b_evol_eq}) is 
(\cite{lich67})
\begin{equation}
u^{\beta} \, {B^{\alpha}}_{; \beta} \; = \; B^{\beta} \, {u^{\alpha}}_{; \beta} 
\; - \; B^{\alpha} \, {u^{\beta}}_{; \beta}
\end{equation}
which uses the standard MTW conventions (Einstein summation, semicolon to signify 
covariant differentiation, and comma for ordinary differentiation).  Expanding the 
covariant derivatives and solving for evolution of $B^{\alpha}$
\begin{eqnarray}
\label{rel_b_evol_eq}
u^{\beta} \, {B^{\alpha}}_{, \beta} & = & B^{\beta} \, {u^{\alpha}}_{, \beta} 
\; + \; B^{\beta} \, u^{\gamma} \, 
( { \Gamma^{\alpha}}_{\gamma \beta} \, - \, {\Gamma^{\alpha}}_{\beta \gamma} )
\; - \; B^{\alpha} \, {u^{\beta}}_{; \beta}
\nonumber
\\
& = & B^{\beta} \, {u^{\alpha}}_{, \beta} \; + \; B^{\beta} \, u^{\gamma} \, 
{c_{\beta \gamma}}^{\alpha}
\; - \; B^{\alpha} \, {u^{\beta}}_{; \beta}
\end{eqnarray}
where the ${\Gamma^{\alpha}}_{\beta \gamma}$ are the connection coefficients and 
the antisymmetric commutation coefficients are given by the components of 
basis vector commutators
\begin{equation}
\label{gen_metric_shear_eq}
{c_{\beta \gamma}}^{\alpha} 
\; \equiv \; \left[ {\bm e}_{\beta}, {\bm e}_{\gamma} \right]^{\alpha}
\end{equation}
The metric shear ${c_{\beta \gamma}}^{\alpha}$ vanishes for $B^{\alpha}$ expressed in a 
coordinate basis.  But it does not for a physical, locally-Lorentz, orthonormal frame
(one in which local observers measure physical quantities, and in which 
magnetic field has the units of Gauss for {\em all} components).  In such an 
orthonormal frame, where we denote the coordinates with a ``hat'' accent, equation 
(\ref{gen_metric_shear_eq}) becomes
\begin{equation}
\label{ortho_metric_shear_eq}
{c_{\hat{\beta} \hat{\gamma}}}^{\hat{\alpha}}
\; = \; {{\cal{L}}^{\mu}}_{\hat{\beta}} \, {{\cal{L}}^{\nu}}_{\hat{\gamma} , \mu} \, 
{{\cal{L}}^{\hat{\alpha}}}_{\nu}
\; - \; 
{{\cal{L}}^{\mu}}_{\hat{\gamma}} \, {{\cal{L}}^{\nu}}_{\hat{\beta} , \mu} \, 
{{\cal{L}}^{\hat{\alpha}}}_{\nu}
\end{equation}
where 
${{\cal{L}}^{\mu}}_{\hat{\beta}}$ is a coordinate transformation that 
diagonalizes the metric, converting to a locally-Lorentz system, with inverse 
${{\cal{L}}^{\hat{\beta}}}_{\mu}$ 
\begin{displaymath}
\eta_{\hat{\beta} \hat{\gamma}} \; = 
\; {{\cal{L}}^{\mu}}_{\hat{\beta}} {{\cal{L}}^{\nu}}_{\hat{\gamma}} \, 
g_{\mu \nu}
~~~~~~~~~~
g_{\mu \nu} \; = 
\; {{\cal{L}}^{\hat{\beta}}}_{\mu} {{\cal{L}}^{\hat{\gamma}}}_{\nu} \, 
\eta_{\hat{\beta} \hat{\gamma}} 
\end{displaymath}

To compute ${c_{\hat{\beta} \hat{\gamma}}}^{\hat{\alpha}}$, and therefore the 
evolution of the magnetic field for rotating black hole magnetospheres, we use 
the zero-charge Boyer-Lindquist version of the Kerr metric (MTW) written in the 
following form
\begin{equation}
\label{4d_kerr_metric_eq}
ds^2 \; = \; - \alpha^2 \, dt^2 \; + \; \frac{\rho^2}{\Delta} dr^2 \; + \; \rho^2 \, d\theta^2 \; 
+ \; \varpi^2 \, \left[ d \phi \, - \, \omega \, dt \right]^2
\end{equation}
where the lapse function and angular velocity of frame dragging are given by
\begin{displaymath}
\alpha \; = \; \frac{\rho}{\Sigma} \, \Delta^{1/2}
~~~~~~~~~~
\omega \; = \; \frac{2 a M r} { \Sigma^2}
\end{displaymath}
the various geometric functions are given by 
\begin{eqnarray*}
\Delta \; \equiv \; r^2 \, - \, 2Mr \, + \, a^2 
~~~~~~~~~~ 
\rho^2 \; \equiv \; r^2 \, + \, a^2 \, cos^2 \theta
\\
\Sigma^2 \; \equiv \; (r^2 + a^2)^2 \, - \, a^2 \, \Delta \, sin^2 \theta
~~~~~~~~~~
\varpi \; \equiv \; \frac{\Sigma}{\rho} \, sin \theta
\end{eqnarray*}
$a \, \equiv \, J/M \, = \, j \, M$ is the specific angular momentum, and 
physical quantities are now in geometric units ($c = G = 1$).  
This metric can be diagonalized with many transformations, of course.  We choose 
one that rotates with the same angular velocity $\omega$ as the frame-dragged space. 
This is the so-called Zero Angular Momentum Observer (ZAMO) system of \cite{tpm86}.  
The non-zero elements of this transformation and its inverse are 
\[ \begin{array}{lclclcl}
{{\cal{L}}^{t}}_{\hat{t}} & = & {1} \, / \, \alpha
& ~~~~~~~~~~ &
\\
{{\cal{L}}^{r}}_{\hat{r}} & = & {\Delta^{1/2}} \, / \, {\rho}
& ~~~~~~~~~~ &
{{\cal{L}}^{\theta}}_{\hat{\theta}} & = & {1} \, / \, {\rho}
\\
{{\cal{L}}^{\phi}}_{\hat{t}} & = & \omega \, / \, \alpha
& ~~~~~~~~~~ &
{{\cal{L}}^{\phi}}_{\hat{\phi}} & = & {1} \, / \, \varpi
\\
\\
{{\cal{L}}^{\hat{t}}}_{t} & = & \alpha
& ~~~~~~~~~~ &
\\
{{\cal{L}}^{\hat{r}}}_{r} & = & {\rho} \, / \, {\Delta^{1/2}}
& ~~~~~~~~~~ &
{{\cal{L}}^{\hat{\theta}}}_{\theta} & = & {\rho}
\\
{{\cal{L}}^{\hat{\phi}}}_{t} & = & - \varpi \, \omega
& ~~~~~~~~~~ &
{{\cal{L}}^{\hat{\phi}}}_{\phi} & = & \varpi
\end{array} \]
with the ${\bm e}_{\hat{r}}$ and ${\bm e}_{\hat{\theta}}$ axes parallel to 
${\bm e}_{r}$ and ${\bm e}_{\theta}$, respectively.  

To illustrate that the metric shear acts like a fluid shear flow, consider the 
situation where there is no plasma motion with respect to the orthonormal system, 
so that the fluid four-velocity is given by $u^{\hat{\alpha}} = (-1,0,0,0)$.  
Expressing equation (\ref{rel_b_evol_eq}) in the orthonormal ZAMO frame, the 
evolution of the $\hat{\phi}$ component of the magnetic field, using 
equation (\ref{ortho_metric_shear_eq}), is
\begin{eqnarray}
{B^{\hat{\phi}}}_{, \hat{t}} & = & B^{\hat{t}}      \, {c_{\hat{t}      \hat{t}}}^{\hat{\phi}}
                           \; + \; B^{\hat{r}}      \, {c_{\hat{r}      \hat{t}}}^{\hat{\phi}}
                           \; + \; B^{\hat{\theta}} \, {c_{\hat{\theta} \hat{t}}}^{\hat{\phi}} 
                           \; + \; B^{\hat{\phi}}   \, {c_{\hat{\phi}   \hat{t}}}^{\hat{\phi}} 
\nonumber
\\
\label{4d_bp_evol_eq}
& = & \varpi \, \frac{\Delta^{1/2}}{\alpha \, \rho} 
\left[ B^{\hat{r}} \, \frac{\partial \omega}{\partial r} \; + \; 
B^{\hat{\theta}} \, \frac{1}{\Delta^{1/2}} \, \frac{\partial \omega}{\partial \theta} \right]
\end{eqnarray}
with the time and azimuthal parts vanishing.  
Comparing with equation (\ref{nonrel_bp_evol_eq}), then, we see that there is 
an exact analogy with the classical rotating fluid shear case.  Any poloidal 
magnetic field line penetrating the space near the rotating black hole is 
sheared into azimuthal field at the expense of the rotational energy of the 
black hole, whether or not the magnetic field line actually threads the hole.
This is independent of any shear in the fluid flow itself that may or may not be present.

Equation (\ref{4d_bp_evol_eq}) also can be derived using the ``3+1'' language of 
\cite{tpm86} (TPM).  In that case the metric equation (\ref{4d_kerr_metric_eq}) is 
expressed in terms of the lapse and shift functions and a 3-metric
\begin{equation}
\label{3p1_kerr_metric_eq}
ds^2 \; = \; - \alpha^2 dt^2 \; + \; \gamma_{ij} \, (dx^i \, + \, \beta^i dt) 
(dx^j \, + \, \beta^j dt)
\end{equation}
where $\gamma_{ij}$ raises or lowers indices on the shift 3-vector $\beta^i$ 
and 3-velocity $v^i$.  Comparing (\ref{4d_kerr_metric_eq}) and (\ref{3p1_kerr_metric_eq}) 
gives $\beta^i = (0, 0, -\omega)$, and components of (the diagonal) $\gamma_{ij}$ 
can be determined similarly.  
Physical quantities like magnetic field and velocity vectors are measured in 
a reference frame that rotates with the ZAMOs, but they and the gradient 
operator ${\bm \nabla}$ are expressed in terms of absolute (Boyer-Lindquist) spatial 
and time coordinates.  Under these conditions, the general relativistic analog to 
equation (\ref{nonrel_b_evol_eq}) is (TPM; \cite{kc96a}) 
\begin{equation}
\label{3p1_b_evol_eq}
\frac{\partial {\bm B}}{\partial t} \, - \, \pounds_{\beta} {\bm B} \, + 
\, \alpha {\bm v}{\bm \cdot}{\bm \nabla}{\bm B} 
\; = \; {\bm B}{\bm \cdot}{\bm \nabla} ( \alpha {\bm v}) \; - \; 
{\bm B}\, {\bm \nabla}{\bm \cdot} ( \alpha {\bm v})
\end{equation}
where $\pounds_{\beta}$ represents the Lie derivative along ${\bm \beta}$ 
\begin{equation}
\label{lie_der_eq}
\pounds_{\beta} {\bm B} \; = \; {\bm \beta} \cdot {\bm \nabla} {\bm B} \; - \; 
{\bm B} \cdot {\bm \nabla} {\bm \beta}
\end{equation}
and the frozen-field condition ${\bm E} = {\bm -v} \times {\bm B}$ and 
solenoidal condition ${\bm \nabla} \cdot {\bm B} = 0$ have been applied.  
We again consider the situation where there is no fluid motion with respect to 
the rotating space (${\bm v} = 0$).  Furthermore, because the gradients now are 
expressed in a coordinate frame, the shear terms in equation (\ref{lie_der_eq}) 
vanish, and the Lie derivative becomes simply
$\pounds_{\beta} B^i \, = \, \beta^j {B^i}_{, j} - B^j {\beta^i}_{, j}$.  
As a result, the azimuthal component of equation (\ref{3p1_b_evol_eq}) becomes, 
in Boyer-Lindquist (un-hatted) coordinates, 
\begin{equation}
\label{3p1_bp_evol_eq}
{B^{\phi}}_{, t} \; - \; \beta^{\phi} \, {B^{\phi}}_{, \phi} \; = \; 
B^j \, \omega_{, j}
\end{equation}
where $j = (r, \theta)$ only, since 
$\partial \beta / \partial \phi \, = \, - \partial \omega / \partial \phi \, = \, 0$.  
The left-hand side is proportional to the time derivative in the ZAMO frame 
($= \alpha {B^{\hat{\phi}}}_{, \hat{t}} / \varpi$), so
transformation of equation (\ref{3p1_bp_evol_eq}) to the locally-Lorentz orthonormal 
coordinates yields an equation identical to (\ref{4d_bp_evol_eq}). 
That is, even though the 3+1 formalism is expressed in the Boyer-Lindquist coordinate 
frame, the metric shear still is present;  but that shear is now embodied in the shift 
vector ${\bm \beta}$, rather than in the commutators ${c_{\beta \gamma}}^{\alpha}$.

Since the sheared space near a Kerr black hole is so similar in nature to that of a 
sheared accretion disk ({\it i.e.}, rotating shear flow in a gravitational field), 
it is likely that an instability similar to the Balbus-Hawley one also is operating.  
Thus, there should be not only an omega-effect enhancing the azimuthal magnetic field, 
but also an alpha-effect, due to the magneto-rotational-generated turbulence, 
enhancing the poloidal field as well.  Since this process draws its energy from 
the hole's rotation, the strength of the magnetosphere is expected to continue to grow 
long after all the accretion disk's angular momentum has been removed, and to levels far 
greater than could be supported by the accreting matter alone.  

\end{appendix}

%
%

%



\begin{figure}
\centering
\vskip 1.0cm
\caption{
\label{sim_fig} 
Results from 13 MHD simulations that are similar to those in MEGPL, but 
with angular velocity of the injected material used as the independent 
parameter rather than poloidal magnetic field strength.  Left panel: 
jet velocity (filled squares) and Mach number (open circles) {\it vs.} 
angular velocity, normalized to the Keplerian value.  Right panel:  
power output of the disk system in the form of jet kinetic power 
(filled squares), total kinetic power (wind plus jet, open diamonds), 
and total power in all forms (filled circles).  
For small values of $\Omega/\Omega_{K}$, the disk power is dominated 
by advected magnetic, thermal, and kinetic energy flowing into the 
corona (whose sum varies little or not at all with $\Omega$). 
For large values of $\Omega/\Omega_{K}$, the disk's magneto-rotational 
power (equation \ref{nonrel_mhd_power_eq}) dominates.
The predicted value of 
$\Omega_{crit}$ (equation \ref{omega_crit_eq}) has been modified to 
account for the 
dynamical adjustment in conditions that occurs in the injection region 
when $\Omega \neq \Omega_{K}$ (see text).  When $\Omega < \Omega_{crit}$, 
the total power is dominated by advected magnetic energy, 
and the kinetic energy is dominated by that in the 
loosely-collimated wind.  As $\Omega$ exceeds $\Omega_{crit}$, the 
velocity and Mach number increase by an order of magnitude or more 
and the jet kinetic power becomes a substantial fraction of the 
total.  In these non-relativistic simulations, eventually rotation 
becomes so fast that the flow de-collimates into a mostly radial wind.}
\end{figure}

\begin{figure}
\centering
\vskip 1.0cm
\caption{
\label{roplane_fig} 
The theoretical radio-optical plane for sub-Eddington accreting radio 
galaxies (left panel, $\dot{m} = 10^{-3}$) and near-Eddington accreting 
quasars (right panel, $\dot{m} = 1.0$).  For galaxies, the optical 
luminosity scales roughly with the galaxy mass, as does the black hole mass.  
For quasars, the optical luminosity scales linearly with the black hole mass.
In both cases, the radio power scales with the square of the normalized black 
hole spin $j$.  Note that only the extended (unbeamed) radio power is considered
here; orientation effects are ignored in these diagrams.  
The upper solid lines show equation (\ref{pmhd_eq}) for a maximal Kerr 
hole.  Middle solid lines show the theoretical FR I/FR II break (equation 
\ref{th_frbr_eq}), which agrees well with the observed OLWL relation for 
radio galaxies ({\it cf.} equations \ref{obs_olwl_eq} and \ref{mod_olwl_eq}). 
Lower solid lines show one measure of radio quietness (equation \ref{pmin_eq}), 
below which the Keplerian disk MHD power dominates the ergospheric disk MHD power.  
Note the much smaller region spanned by FR I quasars than by FR I galaxies. 
FR I sources should comprise a much smaller fraction of quasars than of radio 
galaxies.  
The right panel also shows another estimate of radio quietness 
that makes the FR I quasar region even narrower --- dashed 
lines where the radio to optical/UV flux ratio (1 GHz to 1 PHz) attains 
values of 10 and 100.  
The optical/UV flux was estimated by integrating the power output 
of a standard Shakura \& Sunyaev accretion disk outward from the radius where
the disk achieves an effective temperature of $10^{5} \, \rm{K}$. 
The vertical dot-dashed line shows roughly where the optical/UV luminosity 
is equivalent to the stellar luminosity of a typical spiral galaxy, 
dividing Seyfert galaxies from quasars.  
As discussed in the text, the great majority of radio loud 
Seyferts are expected to occur only at high redshifts.  
}
\end{figure}

%


\begin{thebibliography}{}


\bibitem[Bicknell 1985]{bick85}
    Bicknell, G. V. 1985, Proc. Astr. Soc. Austr., 6, 130. 

\bibitem[Bicknell 1995]{bick95}
    Bicknell, G. V. 1995, \apjs, 101, 29. 

\bibitem[Blandford \& Payne (1982)]{bp82}
    Blandford, R. D. \& Payne, D. G. 1982, \mnras, 199, 883 (BP). 

\bibitem[Blandford \& Znajek (1977)]{bz77}
    Blandford, R. D. \& Znajek, R. 1977, \mnras, 179, 433 (BZ). 

\bibitem[Brandenburg 1996]{brand96}
    Brandenburg, A. 1996, \apjlett, 465, L115. 

\bibitem[Brandenburg et al. 1996]{bnst96}
    Brandenburg, A., Nordlund, A., Stein, R.F. \& Torkelsson, U. 1996, \apjlett,
    458, L45. 

\bibitem[Baum, Zirbel, \& O'Dea 1995]{baum95}
    Baum, S. A., Zirbel, E. L., \& O'Dea, C. P. 1995, \apj, 451, 88. 

\bibitem[Camenzind 1989]{c89}
    Camenzind, M. 1989, in Accretion Disks and Magnetic Fields in 
    Astrophysics, ed. G. Belevedere, (Dordrecht: Kluwer), p. 129. 

\bibitem[Cowling 1934]{c34}
    Cowling, T. G. 1934, \mnras, 94, 39.

\bibitem[DeYoung 1976]{dey76}
    DeYoung, D. 1976, Ann. Rev. Astr. Ap., 14, 447. 

\bibitem[Faber et al 1987]{faber87}
    Faber, S. M. et al. 1987, in Nearly Normal Galaxies, proc. 8th Santa 
    Cruz Summer Wkshp on Astron. \& Astrophys., (New York: Springer-Verlag), 
    P. 175. 

\bibitem[Fanaroff \& Riley 1974]{fr74}
    Fanaroff, B. L. \& Riley, J. M. 1974, \mnras, 164, 31P. 

\bibitem[Hayashi, Shibata, \& Matsumoto (1996)]{hay96}
    Hayashi, M. R., Shibata, K., \& Matsumoto, R. 1996, \apjlett, 468, 
    L37.  

\bibitem[Henriksen 1989]{h89}
    Henriksen, R. N. 1989, in Accretion Disks and Magnetic Fields in
    Astrophysics, ed. G. Belevedere, (Dordrecht: Kluwer), p. 117. 

\bibitem[Jackson \& Wall 1999]{jw99}
    Jackson, C. A. \& Wall, J. V. 1999, \mnras, 304, 160. 

\bibitem[Kellermann et al. 1989]{kell89}
    Kellermann, K. I. et al. 1989, \aj, 98, 1195. 

\bibitem[Khanna \& Camenzind (1996a)]{kc96a}
    Khanna, R. \& Camenzind, M. 1996a, \aap, 307, 665. 

\bibitem[Khanna \& Camenzind 1996b]{kc96b}
    Khanna, R. \& Camenzind, M. 1996b, \aap, 313, 1028. 

\bibitem[Koide et al. 1998a]{koide98a}
    Koide, S., Shibata, K., \& Kudoh, T. 1998a, \apjlett, 495, L63. 

\bibitem[Koide et al. 1998b]{koide98b}
    Koide, S., Shibata, K., \& Kudoh, T. 1998b, \apj, submitted.

\bibitem[Kormendy \& Richstone 1995]{kr95}
    Kormendy, J. \& Richstone, D. 1995, Ann. Rev. Astr. Ap., 33, 581. 

\bibitem[Kudoh \& Shibata 1996]{ks96}
    Kudoh, T. \& Shibata, K. 1996, \apjlett, 452, L41. 

\bibitem[Laing 1996]{laing96}
    Laing, R. A. 1996, in Energy Transport in Radio Galaxies and Quasars,
    eds. P. E. Hardee, A. H. Bridle, and J. A. Zensus, ASP Conf. Series Vol. 
    100, p. 241. 

\bibitem[Leahy 1991]{leahy91}
    Leahy, J. P. 1991, in Beams and Jets in Astrophysics, ed. P. Hughes,
    (Cambridge: Cambridge Univ. Press), p. 100.

\bibitem[Ledlow \& Owen (1996)]{lo96}
    Ledlow, M. J. \& Owen, F. N. 1996, \aj, 112, 9. 

\bibitem[Lichnerowicz 1967]{lich67}
    Lichnerowicz, A. 1967, Relativistic Hydrodynamics and Magnetohydrodynamics; 
    Lectures on the Existence of Solutions, (New York: Benjamin).

\bibitem[Livio, Ogilvie, \& Pringle (1999)]{lop99}
    Livio, M., Ogilvie, G. I., \& Pringle, J. E. 1999, \apj, 512, 100.

\bibitem[MacDonald \& Thorne (1982)]{mt82}
    Macdonald, D. \& Thorne, K. S. 1982, \mnras, 198, 345. 

\bibitem[Meier 1976]{m76}
    Meier, D. L. 1976, \apj, 207, 343. 

\bibitem[Meier 1982]{m82}
    Meier, D. L. 1982, \apj, 256, 681.

\bibitem[Meier, Payne, \& Lind 1996]{mpl96}
    Meier, D. L., Payne, D. G., \& Lind, K. R. 1996, in IAU Symposium 175: 
    Extragalactic Radio Sources, eds. R. Ekers, C. Fanti, and L. Padrielli, 
    (Dordrecht: Kluwer), p. 433. 

\bibitem[Meier 1996]{m96}
    Meier, D. L. 1996, \apj, 459, 185. 

\bibitem[Meier et al. (1997a)]{m97a}
    Meier, D. L., Edgington, S., Godon, P., Payne, D. G., \& Lind, K. R., 
    1997a, Nature, 388, 350. (MEGPL) 

\bibitem[Meier et al. 1997b]{m97b}
    Meier, D. L. et al. 1997b, in IAU Colloquium 164: Radio Emission 
    from Galactic and Extragalactic Compact Sources, eds. J. A. Zensus, 
    G. B. Taylor, \& J. M.  Wrobel, ASP Conf. Series Vol. 144, p. 51. 

\bibitem[Misner, Thorne, \& Wheeler (1973)]{mtw73} 
    Misner, C. W., Thorne, K. S., \& Wheeler, J. A. 1973, {\it 
    Gravitation} (San Francisco: Freeman) (MTW). 

\bibitem[Moderski \& Sikora 1996]{ms96}
    Moderski, R. \& Sikora, M. 1996, \mnras, 283, 854. 

\bibitem[Narayan \& Yi 1995]{ny95}
    Narayan, R. \& Yi, I. 1995, \apj, 444, 231. 

\bibitem[Natarajan \& Pringle 1998]{np98}
    Natarajan, P. \& Pringle, J. E. 1998, LANL astro-ph preprint no. 9808187.

\bibitem[Novikov \& Thorne 1973]{nt73}
    Novikov, I. D. \& Thorne, K. S. 1973, in Black Holes, eds. C. DeWitt 
    \& B. DeWitt, (New York: Gordon \& Breach), p. 343. 

\bibitem[N\'{u}\~{n}ez (1997)]{n97}
    N\'{u}\~{n}ez, M. 1997, Phys. Rev. Lett, 79, 796.

\bibitem[Owen \& Laing (1989)]{ol89}
    Owen, F. N. \& Laing, R. A. 1989, \mnras, 238, 357. 

\bibitem[Owen \& White (1991)]{ow91}
    Owen, F. N. \& White, R. A. 1991, \mnras, 249, 164. 

\bibitem[Punsly \& Coroniti (1990a)]{pc90a}
    Punsly, B. \& Coroniti, F. V. 1990a, \apj, 350, 518. 

\bibitem[Punsly \& Coroniti 1990b]{pc90b}
    Punsly, B. \& Coroniti, F. V. 1990b, \apj, 354, 583 (PC). 

\bibitem[Rees 1978]{rees78}
    Rees, M. J. 1978, Nature, 275, 516. 

\bibitem[Robson 1992]{rob92}
    Robson, I. 1992, in Variability of Blazars, eds. E. Valtaoja and 
    M. Valtonen, (Cambridge: Cambridge Univ. Press), p. 111. 

\bibitem[Romanova et al. (1998)]{rom98}
    Romanova, M. M. et al. 1998, \apj, 500, 703. 

\bibitem[Roberts 1993]{r93}
    Roberts, P. H. 1993, in Astrophysical Fluid Dynamics, eds. J.-P. Zahn
    \& J. Zinn-Justin, (Amsterdam: Elsevier), p. 229. 

\bibitem[Schmidt 1972]{schm72}
    Schmidt, M. 1972, \apj 176, 289. 

\bibitem[Shakura \& Sunyaev 1973]{ss73}
    Shakura, N. I. \& Sunyaev, R. A. 1973, \aap, 24, 337.  

\bibitem[Sobczak et al. 1998]{sobc98}
    Sobczak, G. J., McClintock, J. E., Remillard, R. A., Bailyn, C. D., \&
    Orosz, J. A. 199, \apj, in press.

\bibitem[Stone et al. 1996]{stone96}
    Stone, J.M., Hawley, J.F., Gammie, C.F., \& Balbus, S.A. 1996, \apj, 
    463, 656. 

\bibitem[Tavani et al. 1996]{tav96}
    Tavani, M. et al 1996, \apjlett, 473, L103. 

\bibitem[Thorne, Price, \& MacDonald (1986)]{tpm86}
    Thorne, K. S., Price, R. H., \& MacDonald, D. A. 1986, Black Holes:
    The Membrane Paradigm, (New Haven: Yale Univ. Press).

\bibitem[Tingay et al. 1998]{ting98}
    Tingay, S. J. et al. 1998, \aj, 115, 960. 

\bibitem[Urry \& Padovani 1995]{up95}
    Urry, C. M. \& Padovani, P. 1995, \pasp, 107, 803.

\bibitem[Ustyugova et al. 1999]{ust99}
    Ustyugova et al. 1999, \apj, 516, 221.

\bibitem[Vermeulen 1996]{verm96}
    Vermeulen, R. C. 1996, in Energy Transport in Radio Galaxies and Quasars,
    eds. P. E. Hardee, A. H. Bridle, and J. A. Zensus, ASP Conf. Series Vol. 
    100, p. 117. 

\bibitem[Wall \& Jackson 1997]{wj97}
    Wall, J. V. \& Jackson, C. A. 1997, \mnras, 290, L17. 

\bibitem[Wilson \& Colbert 1995]{wc95}
    Wilson, A. S. \& Colbert, E. J. M. 1995, \apj, 438, 62. 

\bibitem[Windhorst et al. 1995]{wind95}
    Windhorst, R. A. et al 1995, Nature, 375, 471. 

\bibitem[Zhang et al. 1997]{zhang97}
    Zhang, S. N., Cui, W., \& Chen, W. 1997, \apjlett, 482, L155.

\end{thebibliography}
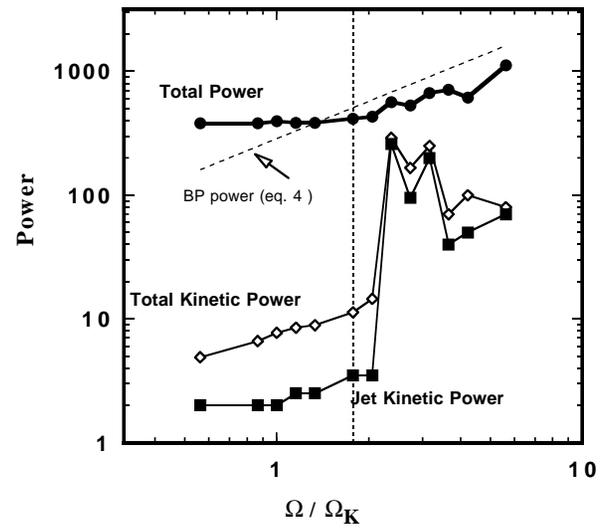
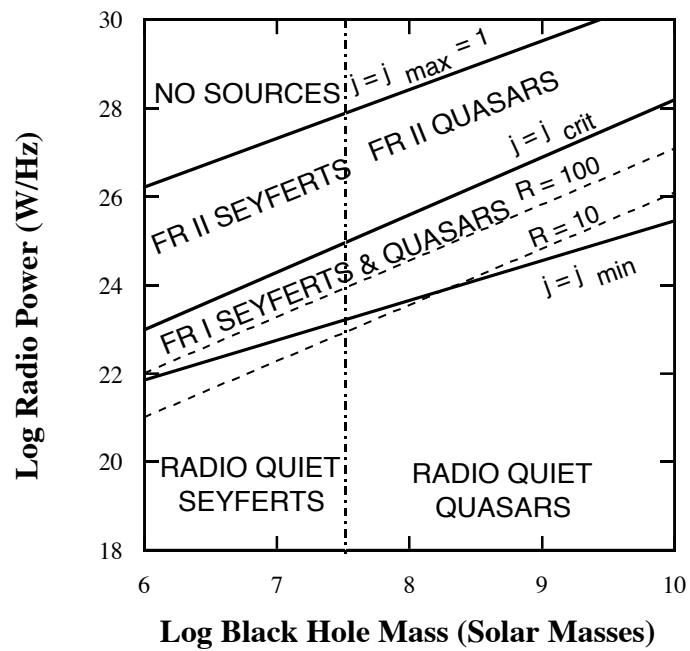
\end{document}